\documentclass[a4paper,aps,prl,twocolumn,amsmath,amssymb,showpacs,floatfix,10pt]{revtex4}
\usepackage{epsfig}
\usepackage{graphicx}
\usepackage{amsmath,amssymb,amsfonts,latexsym}
\newcommand{\be}{\begin{equation}}
\newcommand{\ee}{\end{equation}}

\newcommand{\A}{{\bf A}^{\alpha}({\bf r})}
\newcommand{\B}{{\bf B}^{\alpha}({\bf r})}
\newcommand{\Bpar}{B^{\alpha}_{\parallel}({\bf r})}
\newcommand{\Bperp}{{\bf B}^{\alpha}_{\perp}({\bf r})}

\begin{document}
\title{Critical Phenomena in a Highly Constrained Classical Spin System:\\
N\'eel Ordering from the Coulomb Phase}
\author{T. S. Pickles, T. E. Saunders, and J. T. Chalker}
\affiliation{Theoretical Physics, University
  of Oxford, 1 Keble Road, Oxford, OX1 3NP, United Kingdom}
\date{\today}
\begin{abstract}
Many classical, geometrically frustrated antiferromagnets
have macroscopically degenerate ground states. In a class of
three-dimensional systems, the set of degenerate ground states
has power-law correlations and is an example
of a Coulomb phase. We investigate N\'eel ordering from such a Coulomb
phase, induced by weak additional interactions that lift the
degeneracy. We show that the critical point belongs to a universality
class 
that is 
different from the one for the equivalent transition out of the
paramagnetic phase, and 
that it is 
characterised by effective long-range
interactions; 
alternatively, ordering may be discontinuous.
We suggest that a transition of this type may be
realised by applying uniaxial stress to a pyrochlore antiferromagnet.
\end{abstract}

\pacs{
64.60.Cn        %Order-disorder transformations; stat mech of model systems
75.10.Hk 	%Classical spin models
05.70.Jk        %Critical point phenomena
}

\maketitle

%Paragraph 1: Introduction. GFAFMs, ice, and close-packed dimer models -
%ground state degeneracy and dipolar correlations.
%
%Sets of states that satisfy local constraints yet have 
%extensive degeneracy are of enduring interest in statistical physics.
%Examples include proton configurations in ice \cite{ice}, ground states of
%highly frustrated magnets \cite{gfafm-review}, and dimer covering of
%lattices \cite{dimer-review}. One might imagine that correlation
%functions, averaged over states of this kind, would be short-ranged
%as a consequency of the degeneracy. Strikingly, this need not be the case.

The statistical physics of systems that combine local constraints with
macroscopic degeneracy has been of enduring interest in several
different contexts. Examples include models for geometrically
frustrated antiferromagnets \cite{gfafm-review} and for proton
disorder in ice \cite{ice}, and dimer covering problems on a variety
of lattices \cite{dimer-review}. Local constraints arise from energy
minimisation when ground states of a Hamiltonian are
considered, or are rules that must be satisfied by all allowed configurations, as for
closed-packed dimers on a lattice. Macroscopic degeneracy appears
if there are extensively many ways to satisfy these constraints.
Despite the degeneracy, correlation
functions, averaged over states of this kind, need not be short-range.
Indeed, it has recently been appreciated, both for dimer
coverings of three-dimensional bipartite lattices, and for ground
states of certain geometrically frustrated magnets, that pair correlations
at large separation have a characteristic, algebraic dipolar form
\cite{huse2003,isakov2004,henley2005}. This behaviour is captured by
the Coulomb phase of a theory in which the local constraints
of lattice models become the requirement that a continuum field is
solenoidal. The Coulomb phase is stable against all small perturbations that
are allowed by symmetry, but a sufficiently large perturbation may induce
long range order.

In this paper we consider a N\'eel ordering transition that takes place
from the Coulomb phase of a geometrically frustrated antiferromagnet,
asking how it differs from the conventional N\'eel transition out of
the paramagnetic phase. Our interest in such a transition stems in
part from the fact that, by design, it evades a naive application of
Landau-Ginzburg-Wilson theory, since a description based in the standard way
on symmetries and a local order parameter for the 
N\'eel phase will not reproduce the
correlations of the Coulomb phase \cite{bergman2006}. Similar considerations were also a motivation
for an intriguing recent study of another ordering transition from a Coulomb
phase, involving close-packed dimers on a cubic lattice \cite{alet2006}. In
that case, high-precision Monte Carlo simulations indicate critical
behaviour different from that expected at a conventional critical
point with the same symmetries, but an analytical
understanding of the transition has yet to be developed. 
By contrast, we show in the following how the N\'eel ordering
transition from the Coulomb phase 
may be brought within the standard theoretical
framework used to describe critical phenomena.
More broadly, in light of the intense current interest in quantum
phase transitions that lie beyond the scope of a conventional Landau
treatment \cite{science},
%,motrunich2005,bergman2006
it is natural to
examine ordering from the Coulomb phase as a classical
counterpart.
%%%%%%%%%%%%%%%%%%%%%%%%%%%[1]%%%%%%%%%%%%%%%%%%%%%%%%%%%%%%%%%%%%%%%%%%%%%%%
\begin{figure}[ht]
\begin{center}
\includegraphics[width=6.5cm]{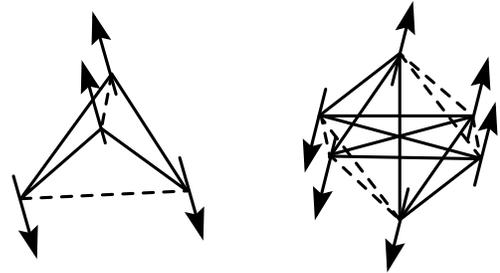}
\caption{N\'eel order induced by uniaxial compression in the building
  blocks of geometrically frustrated antiferromagnets. Left: a
  tetrahedron from the
  pyrochlore lattice strained along $[100]$. Right: an octahedron from
  a lattice of corner-sharing octahedra strained along $[111]$. Full
  lines denote antiferromagnetic exchange interactions of strength
  $J$, and dashed lines indicate ones of strength $J-\Delta$. Arrows
  represent spins. 
}
 \label{distorted-units}
\end{center}
\vspace{-0.5cm}
\end{figure}
%%%%%%%%%%%%%%%%%%%%%%%%%%%%%%%%%%%%%%%%%%%%%%%%%%%%%%%%%%%%%%%%%%%%%%%%

The class of models we are concerned with is as follows. We study 
classical antiferromagnets on geometrically frustrated,
three-dimensional lattices that are known to support a Coulomb phase in the zero
temperature limit when all non-zero exchange interactions are the same. 
These systems meet three requirements: they have
macroscopically degenerate ground states, do not
display order by disorder, and have a bisimplex lattice.
Such lattices consist of corner-sharing frustrated units,
or simplices, which are arranged in a bipartite fashion, with
exchange interactions linking all pairs of spins within each simplex. 
We focus on two examples: the Heisenberg model on the 
pyrochlore lattice, built from corner-sharing tetrahedra, 
which is approximately realised in a number of materials; and a system of
$n$-component spins on a lattice of corner-sharing octahedra, which is
a convenient theoretical construct. The local constraints satisfied in
ground states of these models are that the total spin of each simplex
is zero \cite{moessner1998}. We induce order using a
change in the relative strength of exchange interactions which can be
produced by uniaxial compression via magnetoelastic
coupling,  as illustrated in Fig.~\ref{distorted-units}. Exchange
energies $J_{ij}$ within each simplex then take the two values $J$ and $J-\Delta$, and our focus
is the regime $J \gg \Delta > 0$. The Hamiltonian for both models
has the form
\begin{equation}
{\cal H} = \sum_{\langle ij \rangle} J_{ij}\, {\bf S}_i \cdot {\bf S}_j\,
\label{spin-hamiltonian}
\end{equation}
where ${\bf S}_i$ is an $n$-component unit vector.
In each case, positive $\Delta$ removes the
macroscopic ground state degeneracy of the unstrained system,
selecting from these ground states the two-sublattice N\'eel states of
Fig.~\ref{distorted-units}, in which neighbouring pairs of spins coupled by the
stronger interaction $J$ are antiferromagnetically aligned, while
frustration forces those pairs coupled by the weaker interaction $J-\Delta$ 
to have ferromagnetic alignment. Considering the effect of thermal
fluctuations, one expects N\'eel order below
a critical temperature $T_{\rm c} \sim \Delta$, and a disordered phase
at higher temperatures. For temperatures $T \ll J$, thermal
fluctuations are predominantly within the ground
state manifold of the unstrained system, and in the limit $J/\Delta
\rightarrow \infty$ the transition at $T_{\rm c}$ is between a N\'eel
phase and a Coulomb phase. 

Several examples are known of magnetic ordering
associated with lattice strain in frustrated magnets, although
we are not aware of a direct realisation of the transition we discuss.
In particular, N\'eel order has been induced by uniaxial stress
in the pyrochlore $\rm Tb_2 Ti_2 O_7$ \cite{mirebeau}, while it is
accompanied by spontaneous strain in the spinel $\rm Zn Cr_2 O_4$
\cite{chromite}. Unfortunately, $\rm Tb_2 Ti_2 O_7$ is a material that is at present
poorly understood, and the transition in $\rm Zn Cr_2 O_4$ is strongly first order.  
We hope that this paper will stimulate further experiments in the area.

We set out below several complementary approaches to
understanding ordering in constrained systems. Starting
from the continuum description of the Coulomb phase,
N\'eel order amounts to condensation of the solenoidal field. When this
transition is continuous, we show that it has the
same critical behaviour as for an unconstrained system in which there are 
uniaxial dipolar interactions. From classic studies
\cite{larkin,aharony,brezin}, it is known that 
long-range interactions of this type lead to an upper critical
dimension of three, and we confirm this for the lattice models in the
large-$n$ limit. We supplement this analysis with Monte Carlo
simulations. We find that the transition in a Heisenberg model is
first order, but show that a critical point can be reached by applying
a staggered field.

We start by recalling the mapping of Refs.~\onlinecite{isakov2004} and
\onlinecite{henley2005}, from
ground states at $\Delta = 0$ to solenoidal fields. The mapping uses
the underlying simplex lattice for the spin system, which is a diamond
lattice for the pyrochlore antiferromagnet, and a simple cubic
lattice for the octahedral antiferromagnet. In each case, spins 
${\bf  S}_i$ are located at mid-points of bonds $i$ on the simplex
lattice. Unit vectors $\hat{\bf e}_i$ are defined on these bonds,
with the orientation convention that they are all directed 
from a given simplex sublattice towards the other one.
%into the simplices on one sublattice, and out of the ones on the other sublattice.
A spin configuration is represented using  $n$
flavours of a three-component field, defined
from the spin components ${S}^{\alpha}_i$ by ${\bf B}^{\alpha}_i =
S^{\alpha}_i \hat{\bf e}_i$. The ground state constraint, that the total
spin of each simplex is zero, translates into the condition that these
fields have lattice divergence zero. Introducing continuum coordinates ${\bf
  r}$ and coarse-grained fields ${\bf B}^{\alpha}({\bf r})$
satisfying $\nabla \cdot \B = 0$, the entropic weight on ground states is postulated
\cite{isakov2004,henley2005} to be
proportional to $e^{-{\cal H}_0}$, where
\begin{equation}
{\cal H}_0 = \frac{\kappa}{2} \int {\rm d}^3{\bf r} \sum_\alpha |{\bf
  B}^\alpha({\bf r})|^2 
\label{coulomb-hamiltonian}
\end{equation}
The stiffness $\kappa$ is the single parameter of the theory: it
simply controls the amplitude of correlations.

Under the same mapping, the N\'eel states of Fig.~\ref{distorted-units}
are represented by a uniform, non-zero ${\bf B}^\alpha({\bf r})$. The
orientation of this field
in real space is along $[100]$ for the pyrochlore antiferromagnet, and
along $[111]$ for the octahedral antiferromagnet, while the
orientation in flavour space reflects the ${\rm O}(n)$ symmetry-breaking of
the N\'eel order. 

In order to describe a phase transition between the Coulomb
and N\'eel phases, we incorporate the effect of non-zero $\Delta$ into the continuum
theory of Eq.~(\ref{coulomb-hamiltonian}) via a temperature-dependent
control parameter $t$, with $t=1$ for $T\gg \Delta$ and $t<0$ for $T
\ll \Delta$. Separating ${\bf B}^{\alpha}({\bf
  r})$ into components $B^{\alpha}_{\parallel}({\bf r})$  and  
${\bf B}^{\alpha}_{\perp}({\bf r})$, respectively parallel and perpendicular
to the preferred direction, we replace ${\cal H}_0$ 
by ${\cal H}_2 + {\cal H}_4$, where
\begin{eqnarray}
{\cal H}_2 = \frac{\kappa}{2} \int {\rm d}^3{\bf r} \sum_{\alpha} \left\{ 
|\Bperp|^2 + t |\Bpar|^2 + |{\bf \nabla} \times \B|^2
\right\} \nonumber \\
{\cal H}_4 = u \int {\rm d}^3{\bf r}\big\{\sum_{\alpha} |\B|^2\big\}^2\,.\qquad \qquad \qquad
\label{H}
\end{eqnarray}
At $t=1$, Eq.~(\ref{H}) is simply ${\cal H}_0$ supplemented by
gradient and quartic terms that are irrelevant in the scaling sense
in the Coulomb phase. On reducing $t$
a transition occurs (at $t=0$ within mean field theory) to an ordered phase in which
$\langle \Bpar \rangle \not= 0$. The quartic term ${\cal H}_4$ is
required for stability in the ordered phase, while the gradient term
is necessary to suppress short-wavelength fluctuations close to the critical point.

The solenoidal constraint can be re-expressed as a long-range
interaction. To this end,
write the fields $\B$ in terms of vector potentials $\A$ by setting $\B
= \nabla \times \A$, and choose the Coulomb gauge $\nabla\cdot\A = 0$.
Then ${\cal H}_2$ is diagonalised by Fourier transform and by rotation in the 
space of the three components of ${\bf A}^{\alpha}$. In a fixed gauge
only two of these components fluctuate independently: taking their
amplitudes at wavevector ${\bf q}$ to be $a^\alpha_1({\bf q})$ and
$a^\alpha_2({\bf q})$, the diagonalised form is
%\begin{equation}
%\Aq = \int d^3{\bf r} \A e^{i {\bf q}.{\bf r}}\,.
%\end{equation}
%Then we find (taking $L$ large)
%\begin{equation}
%\sum_a \int d{\bf r} \left\{ 
%|\Bperp|^2 + t |\Bpar|^2 \right\}
%=\frac{1}{(2 \pi)^3}\int d{\bf q}\sum_{a}\sum_{ij} A_i(-q) M_{ij}(q) A_j(q)\,.
%\end{equation}
%Choosing $q_{\parallel}$ to lie along the $x$-direction, the matrix
%${\bf M}({\bf q})$ has the form
%\begin{equation}
%\left(
%\begin{array}{ccc}
%q_y^2 + q_z^2 & -q_x q_y       & -q_x q_z\\
%-q_x q_y      & q_x^2 +t q_z^2 & -t q_y q_z\\
%-q_x q_z      & -t q_y q_z     & q_x^2 +t q_y^2
%\end{array}
%\right)
%\end{equation}
%The eigenvalues $\mu_l({\bf q})$ and eigenvectors ${\bf v}_l({\bf q})$
%of this matrix are
%\begin{equation}
%\begin{array}{ll}
%\mu_1({\bf q}) = q_x^2 + t[q_y^2 + q_z^2] \qquad  &
%{\bf v}_1^{\rm  T}({\bf q}) = c_1(0, - q_z, q_y) \\
%\mu_2({\bf q}) = q^2    &
%{\bf v}_2^{\rm  T}({\bf q}) = c_2(-[q_y^2 + q_z^2], q_x q_y, q_x q_z)\\
%\mu_3({\bf q}) = 0  &
%{\bf v}_3^{\rm  T}({\bf q}) = c_3(q_x, q_y, q_z)
%\end{array}
%\end{equation}
%where the $c_l$ are normalisation coefficients.  Expanding the vector potential in this basis, with
%$$
%{\bf A}^a({\bf q}) = \sum_l a^a_l({\bf q}) {\bf v}_l({\bf q})\,,
%$$
%the condition $\nabla \cdot \A = 0$ is equivalent to $a^a_3({\bf q})=0$. In
%addition, as the transition is approached, only $\mu_1({\bf q})$ and
%not $\mu_2({\bf q})$ softens, and so we expect the critical field to
%be $a_1({\bf q})$.
%We have 
\begin{eqnarray}
{\cal H}_2 = \frac{\kappa}{2} \int d^3{\bf q} \sum_{\alpha} && 
\Big\{
[q_{\parallel}^2 + t q_{\perp}^2 + q^4] a^{\alpha}_1(-{\bf q}) a^{\alpha}_1({\bf
q}) \nonumber\\ && 
+ [q^2 + q^4] a^{\alpha}_2(-{\bf q}) a^{\alpha}_2({\bf
q}) \Big\}\,.
%\nonumber\\&+& u \int d^3{\bf q}_1 \int d^3{\bf q}_2 \int d^3{\bf q}_3
%\,\,\sum_{a b}a^a_1({\bf q}_1) a^a_1({\bf  q_2}) a^b_1({\bf q}_3) a^b_1(-{\bf q}_1 -{\bf q}_2 -{\bf q}_3)\,.
\end{eqnarray} 
To emphasise the equivalence to a system with uniaxial
dipolar interactions, we  rescale the fields by a factor $q$, writing
$\varphi_l^{\alpha}({\bf q}) = q a_l^{\alpha}({\bf q})$ for $l=1,2$; and since
$\varphi_2^{\alpha}({\bf q})$ is non-critical, we omit it, obtaining finally
\begin{eqnarray}
{\cal H}_2 &=& \frac{\kappa}{2} \int d^3{\bf q} \sum_\alpha 
\left[(1-t)\frac{q_{\parallel}^2}{q^2} + t  + q^2\right] \varphi^\alpha_1(-{\bf q}) \varphi^\alpha_1({\bf
q}) \,.\nonumber\\
%+ [1 + q^2]\varphi_2(-{\bf q}) \varphi_2({\bf q}) 
%&+& u \int d^3{\bf q}_1 \int d^3{\bf q}_2 \int d^3{\bf q}_3 \,\,\sum_{a b}a^a_1({\bf q}_1) a^a_1({\bf
%  q_2}) a^b_1({\bf q}_3) a^b_1(-{\bf q}_1 -{\bf q}_2 -{\bf q}_3)\,,
\label{dipolar}
\end{eqnarray}  

In this form, ${\cal H}_2$ has been studied extensively as a model for
uniaxial ferroelectrics and ferromagnets, with
the term $q_\parallel^2/q^2$, non-analytic at $q=0$, arising from
dipolar interactions. Ordering in these systems is
celebrated as an example of a phase transition with an upper 
critical dimension of three, at which critical behaviour is almost
mean field, but with logarithmic corrections  \cite{larkin,aharony,brezin}.
%For example, the susceptibility
%conjugate to the order parameter diverges as $\chi \sim |t|^{-1}
%|\ln(t)|^{(n+2)/(n+8)}$ 
%\cite{larkin,aharony}. 
We conclude that the
same behaviour will appear at the N\'eel ordering transitions
of the constrained spin systems discussed here, a demonstration of the
controlling influence that the correlations of the Coulomb phase have
on critical phenomena.

We next turn to lattice models, and first present an alternative way
of viewing the equivalence between the constrained spin systems we study and ones with
uniaxial dipolar interactions. 
%To be definite, we discuss Eq.~(\ref{spin-hamiltonian}) for the
%octahedral lattice; we have also obtained equivalent results for the pyrochlore lattice. 
Consider the the spectrum of the interaction matrix $J_{ij}$ appearing in
Eq.~(\ref{spin-hamiltonian}). It is block-diagonalised by Fourier
transform with wavevector $\bf k$.
In the class of models we consider, at
$\Delta = 0$ its lowest branch is dispersionless, and for bisimplex
lattices this branch is degenerate with dispersive branches
at points ${\bf k}_{\rm D}$ in the Brillouin zone.
%${\bf k}_\pi \equiv (\pi, \pi, \pi)$. Such flat branches are
A natural approach to statistical mechanics at $T \ll J$ is via projection onto
these flat bands. The projection operator $P$ is non-analytic in $\bf k$ at
${\bf k}_{\rm D}$ and as a result is power-law
in real space  at large distances \cite{descloizeaux}. For 
three-dimensional, bisimplex frustrated antiferromagnets, its
long-distance form is dipolar \cite{isakov2005}.

The dipolar nature of $P$, applied to certain pyrochlore magnets
(spin-ice materials \cite{spin-ice-review}) 
turns out to have the remarkable consequence that dipolar
interactions are essentially equivalent to nearest-neighbour ones,
within the set of low-energy states allowed by local constraints, a
feature termed projective equivalence
\cite{isakov2005}. Here we take an opposite path: we show that
projection of $J_{ij}$ for $\Delta > 0$ turns nearest-neighbour
interactions into long-range ones. 

On favouring N\'eel order by setting $\Delta >0$, degeneracy in the
spectrum of $J_{ij}$ is lifted, leaving isolated minima.
In this language, the models of Fig.~\ref{distorted-units}
are engineered so that these minima occur at the degeneracy points
${\bf k}_{\rm D}$. Then for $\Delta \ll J$ the interaction matrix
inherits the singularities of the projection operator.
To be definite, we now focus on the octahedral lattice, for which
${\bf k}_{\rm D} = (\pi, \pi, \pi)$; we have
also obtained equivalent results for the pyrochlore lattice.
Introducing 
${\bf q} = {\bf k} - {\bf k}_\pi$, the
projected interaction has a lowest eigenvalue that varies with $\bf q$
as
\begin{equation}
{\cal E}({\bf q}) = {\cal E}_0 + 3\Delta \frac{q_{\parallel}^2 }{q^2}
+ {\cal O}(q^2)\,,
\label{projected-spin-H}
\end{equation}
where $q_{\parallel}$ is the component of $\bf q$ in the $[111]$
direction and ${\cal E}_0$ is a constant. Spin configurations that are
thermally allowed for $T/J \rightarrow 0$ are invariant under
projection: the presence in
Eq.~(\ref{projected-spin-H}) of the singular term
$q_{\parallel}^2/q^2$ is therefore a key result, indicating effective uniaxial dipolar
interactions within these configurations. 

With $T/J$ small but non-zero, the power-law correlations of the Coulomb
phase are cut off at a large distance $\xi_{\rm C}$. In these
circumstances, uniaxial dipolar interactions will control critical
phenomena provided the correlation length $\xi$ is smaller than
$\xi_{\rm C}$, but with a crossover to conventional behaviour sufficiently
close to $T_{\rm c}$.

To study the phase transition in the lattice model, we use the large-$n$ limit,  
%need to take thermal averages over states allowed by
%local constraints, with a Boltzmann weight derived from the projected
%interaction. This can be done in the limit $n\rightarrow \infty$,
extending the treatment of geometrically frustrated systems
described in \cite{isakov2004,canals2001}, and recovering
the critical behaviour known for uniaxial dipolar systems at
$n\rightarrow \infty$ \cite{larkin,aharony,brezin}.
As a function of reduced temperature $t_{\rm r}$, we find that the
staggered susceptibility $\chi$ diverges as $\chi \sim |t_{\rm r}|^{-1}
\cdot \ln(1/|t_{\rm r}|)$, and the sublattice magnetisation vanishes as $m
\sim |t_{\rm r}|^{1/2}$.
% and (iii) the singular
%part of the heat capacity varies as $C \sim [\ln(|t_{\rm r}|)]^{-3}$.

To examine these ideas further, we use Monte Carlo
simulations. We study the pyrochlore Heisenberg
antiferromagnet, treated extensively at $\Delta=0$ in previous work
\cite{moessner1998}. Crucially, the absence of energy barriers in this
model at $\Delta=0$ makes it possible to equilibrate simulations at
temperatures as low as $T=10^{-4}J$, permitting access to the regime
we are concerned with. We focus on the two choices:
$\Delta=0.1J$ and $\Delta=10^{-3}J$, using systems of $N=4L^3$ spins,
with $N \leq 4000$ ($L \leq 10$) and employing parallel
tempering \cite{parallel}. 

\begin{figure}[htb]
\begin{center}
\includegraphics[width=8.5cm,angle=0]{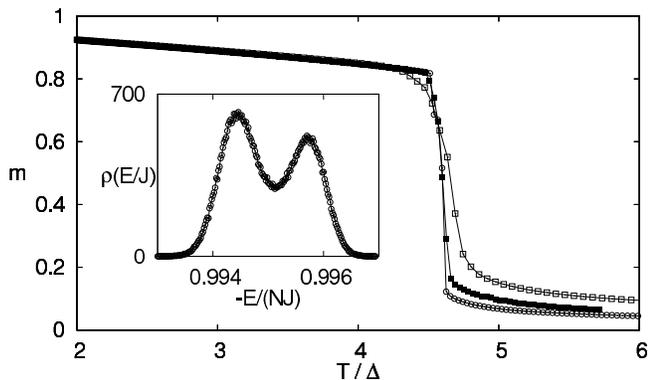}
\caption{Evidence for a first-order transition from simulations with
  $\Delta = 10^{-3}J$. Main panel: $m$ vs. $T/J$ for $L=6$ ($\Box$),
  8 ($\blacksquare$), and 10 ($\circ$). Inset:
  probability distribution $\rho(E/J)$ for the energy $E$ at the 
estimated transition temperature, $T/J=4.6 \times 10^{-3}$, for $L=6$. 
}
 \label{first-order}
\end{center}
%\vspace{-0.5cm}
\end{figure}

For $\Delta = 0.1J$, we find (data not shown) a continuous transition
at $T_c = 0.267J$ and critical
behaviour consistent with the standard, short-range universality class,
which is unsurprising since 
$\xi_{\rm C}$  is not large at this temperature
($\xi_{\rm C} \propto (J/T)^{1/2}$ for $n\geq 2$ and $T\ll J$). For $\Delta=
10^{-3}J$ the data illustrated in Fig.~\ref{first-order} indicate a
strongly first-order transition. The N\'eel order parameter shows a  
rapid change with temperature, and this becomes more step-like
with increasing system size. Moreover, close to the transition
the probability distribution for 
the energy develops two peaks, a feature characteristic of a
first-order transition \cite{energy-distribution}.
%%%%%%%%%%%%%%%%%%%%%%%%%%%[2]%%%%%%%%%%%%%%%%%%%%%%%%%%%%%%%%%%%%%%%%%%%%%%%
%%%%%%%%%%%%%%%%%%%%%%%%%%%%%%%%%%%%%%%%%%%%%%%%%%%%%%%%%%%%%%%%%%%%%%%%

\begin{figure}[t]
\begin{center}
\includegraphics[width=8.5cm,angle=0]{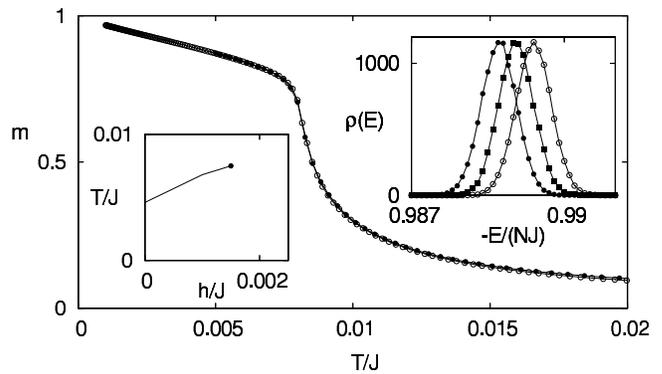}
\caption{Transition in a staggered field with $\Delta =
  10^{-3}J$. Left inset: Schematic 
  phase diagram, showing first-order transition line terminating at a
  critical end point. Main panel: smooth variation of $m$ with $T$ at $h=2\times 10^{-3}J$,
  which is greater than $ h_{\rm c}$, for $L=6$ ($\circ$) and $L=8$
  ($\bullet$). Right inset: probability distribution
  $\rho(E/J)$ for the energy $E$ at
  $T/J=8.0 \times 10^{-3},$ $8.3 \times 10^{-3}$ and $8.6 \times 10^{-3}$, showing single peaks.
}
 \label{continuous}
\end{center}
%\vspace{-0.5cm}
\end{figure}

There is an obvious interest in finding a version of this ordering
transition which is continuous even at $\Delta \ll J$, as a
realisation of the uniaxial critical behaviour we have discussed. 
It is possible that the order of the transition could be altered
%It is possible that the first-order character of the transition shown
%in Fig.~\ref{first-order} reflects a negative fourth-order coefficient
%in a Landau expansion [$u$ in Eq.~(\ref{H})], and so might be altered
by a change in microscopic details of the system. Here, however, we follow
an alternative and more automatic route to a critical point, by applying a staggered
field: we add to Eq.~(\ref{spin-hamiltonian}) the term $-h \sum_i n_i
S_i^z$, with $n_i = \pm 1$ taking opposite signs on
the two N\'eel sublattices. 
%The symmetry-breaking field $h$ induces
%non-zero sublattice magnetisation $m$ at all temperatures. 
For $h \ll \Delta$,
a first-order transition survives, between a high-temperature phase with
small sublattice magnetisation $m$ and a low-temperature one with large $m$. Increasing $h$,
the discontinuity in $m$ at the transition is reduced, and above a
critical field $h_{\rm c}\sim \Delta$ there is no thermally driven phase
transition. The critical end point at $h=h_{\rm c}$ is the continuous
transition we seek. Evidence for its existence is presented in
Fig~\ref{continuous}. 
A precise numerical study of critical behaviour 
at $h=h_{\rm c}$ for $\Delta/J \ll 1$ would be challenging and we do
not attempt it here, but we believe the theoretical case for uniaxial
dipolar critical behaviour at this transition is compelling. 
We note that, while the staggered field used in our simulations cannot be
realised in a Heisenberg system, it would arise from a {\em uniform}
external field in spin-ice materials \cite{spin-ice-review}. 

In summary, we have shown that ordering in highly constrained spin systems
can be quite different from its counterpart in conventional systems. 
We suggest that experiments on ordering in uniaxially strained
geometrically frustrated magnets should be of great 
interest.

We thank L. Balents, R. Moessner, V. Pasquier, and
particularly P. C. W. Holdsworth,
for helpful discussions. This work was supported by EPSRC Grant No.
GR/R83712/01. Part of it was done while JTC was a visitor at KITP Santa
Barbara, supported by NSF Grant No. PHY99-07949.

%%%%%%%%%%%%%%%%%%%%%%%%%%%[3]%%%%%%%%%%%%%%%%%%%%%%%%%%%%%%%%%%%%%%%%%%%%%%%
%%%%%%%%%%%%%%%%%%%%%%%%%%%%%%%%%%%%%%%%%%%%%%%%%%%%%%%%%%%%%%%%%%%%%%%%

%\begin{figure}[t]
%\begin{center}
%\epsfig {figure=figure.eps, width=50mm,angle=270}
%\caption{}
%\label{fig}
%\end{center}
%\vspace{-5mm}
%\end{figure}

\end{document}